
\input harvmac

\noblackbox

\def\a{_{\alpha}}
\def\b{_{\beta}}
\def\eps{\epsilon}
\def\A{\vec A}

\def\T{{k_B T}}

\centerline{\null}
\bigskip
\bigskip
\centerline{\bf Self-Consistent Theory of Normal-to-Superconducting Transition}
\bigskip
\bigskip
\bigskip
\centerline{Leo Radzihovsky}
\centerline{The James Franck Institute and Physics Department}
\centerline{University of Chicago}
\centerline{5640 South Ellis Avenue}
\centerline{Chicago, IL 60637}
\bigskip
\bigskip
\bigskip
\bigskip
\bigskip
\bigskip
\centerline{{\bf ABSTRACT}}
\bigskip
\vbox{
I study the normal-to-superconducting (NS) transition within
the Ginzburg-Landau
(GL) model, taking into account the fluctuations in the $m$-component complex
order parameter $\psi\a$
and the vector potential $\vec A$ in the arbitrary dimension $d$, for any $m$.
I find that the transition is of second-order and that the previous
conclusion of the fluctuation-driven first-order transition is an artifact
of the breakdown of the $\eps$-expansion and the inaccuracy of the
$1/m$-expansion for physical values $\eps=1$, $m=1$. I compute the anomalous
$\eta(d,m)$ exponent at the NS transition, and find $\eta (3,1)\approx-0.38$.
In the $m\rightarrow\infty$ limit, $\eta(d,m)$ becomes exact
and agrees with the $1/m$-expansion.  Near $d=4$ the theory is also
in good agreement with the perturbative $\eps$-expansion results for $m>183$
and provides a sensible interpolation formula for arbitrary $d$ and $m$.}
\vfill\eject

Normal-to-{\it neutral}-superfluid transition is one of the best understood
second-order phase transitions with an unparalleled agreement between theory,
simulations and experiments. In contrast the problem of the normal-to-{\it
charged}-superfluid i.e. the normal-to-superconducting (NS) transition
is significantly more challenging. This problem was first studied
twenty years ago by Halperin, Lubensky and Ma (HLM)
\nref\HLM{B.~I.~Halperin, T.~C.~Lubensky and S.~K.~Ma, {\it Phys.~Rev.~Lett.}
{\bf 32}, 292 (1974).}
\refs{\HLM}
with the GL model generalized
to $m$ complex component $\psi\a$ superconducting order parameter. Using
renormalization group (rg) together with the first order expansion in
$\eps=4-d$
and $1/m$ to treat the gauge field and the order parameter critical
fluctuations,
these authors found that the charge is a relevant operator that grows in the
long wavelength limit. Although for an unphysically large number of order
parameter components, $m>365.9/2\approx 183$, the Heisenberg fixed point
(which controls the neutral superfluid transition)
was found to be unstable to a new critical point at a
finite value of the charge, for physical superconductors ($m=1$) no new
perturbative critical point was found to terminate this charge instability.
The authors interpreted these runaway rg flows as a
signal of a fluctuation-driven first-order phase transition, providing a
first example in which the fluctuations modify the order of the
transition. Similar conclusions were also reached for the scalar
electrodynamics,
exactly in four dimensions in the context of quantum field theory.
\nref\CW{S.~Coleman and E.~Weinberg, {\it Phys.~Rev.~D} {\bf 7}, 1988
(1973).}\refs{\CW}

Although this interpretation is believed to be correct near $d=4$,
the conclusion of the first-order transition for the extreme type-II
superconductors in $d=3$, is most certainly suspect.
Following the original work of Ref.\HLM\ \ Dasgupta and Halperin
\nref\DH{C.~Dasgupta and B.~I.~Halperin, {\it Phys.~Rev.~Lett.}
{\bf 47}, 1556 (1981).}\refs{\DH}
studied the problem on the lattice.  Using duality arguments together
with Monte Carlo techniques, they found that a 3d
superconductor exhibits a second-order transition in the universality
class of the (inverted) XY-model. Later Monte Carlo simulations in 3d further
demonstrated that the nature of the transition changes from first- to
second-order as one goes from the type-I to the extreme type-II
superconductor.
\nref\Bartholomew{J.~Bartholomew, {\it Phys.~Rev.~B} {\bf 28}, 5378
(1983).}\Bartholomew\

In high T$_c$ superconductors the thermal fluctuations are enhanced
and lead to an increase of the critical region by
several orders of magnitude as compared to the conventional superconductors.
Unfortunately, even in these materials, the size of the critical region
is still much too small to experimentally
resolve the question of existence of fluctuation-driven first-order
NS transition, and therefore the nature of the transition appears
to be an academic question.  However, it is believed that
the nematic-to-smectic-A (NA) transition in liquid crystals is
described by a model very similar to the GL gauge theory of the
NS transition,
\nref\deGennes{P.~G.~de Gennes, {\it Solid State Commun.} {\bf 10}, 753
(1972).}\refs{\deGennes}
and therefore the same conclusions apply to this system.
\nref\HL{B.~I.~Halperin and T.~C.~Lubensky,
{\it Solid State Commun.} {\bf 14}, 997 (1974).}\refs{\HL}
In contrast to superconductors, however, the NA transition
is estimated to have a critical region and the size of the
fluctuation-driven first-order transition to be within the
experimentally accessible range.
Since the NA transition appears experimentally
\nref\Experiment{J.~D.~Litster, et al.,
{\it Ordering in Strongly Fluctuating
Condensed Matter Systems}, edited by T.~Riste (Plenum, NY, 1980).}
\refs{\Experiment}, and predicted theoretically
\nref\Toner{J.~Toner, {\it Phys.~Rev.~B} {\bf 26}, 462 (1982).}\Toner\ \
to be continuous, I take this as a further indication of the breakdown of the
perturbative $\eps$-expansion in 3d and question the conclusions
of Ref.\HLM.

In this Letter, I reexamine the problem of the NS transition
with analytical methods that do not rely on the perturbative expansion
in $\eps$ or $1/m$. Using a non-perturbative
method which amounts to solving approximate Dyson equations
for arbitrary $d$ and $m$,
\nref\SCSAmethod{A.~J.~Bray, {\it Phys.~Rev.~Lett.} ${\bf 32}$, 1413 (1974).
Recently the SCSA was applied to a problem of tethered membranes
where it predicted exponents that are exact in $m\rightarrow\infty$,
$m=0$ and correct to the leading order in $\epsilon$;
P. Le Doussal, L. Radzihovsky, {\it Phys.~Rev.~Lett.} ${\bf 69}$ 1209
(1992).}\refs{\SCSAmethod}
I find a nontrivial critical fixed point ($e\neq 0$) that controls the
NS transition. When the order parameter and the gauge field
fluctuations are taken into
account the Heisenberg critical point ($e=0$) controlling the neutral
superfluid transition is found to be unstable to this new critical point.
I therefore show that in contrast to the previous conclusions based on the
$\eps$-expansions, the 3d type-II superconductors undergo a second-order
NS transition, consistent with the consensus described above.
Besides being an independent prediction for the
nature of the NS transition in 3d, corroborating the findings of
Ref.\DH\ \ , my approach has the advantage of working in arbitrary
dimension and therefore sheds light on the question of how the 3d behavior is
connected to the findings near $d=4$.

Within the GL description, the generalized superconductor
is defined by the free-energy functional $F[\psi\a,\A]$ of the
$m$-complex-component superconducting order parameter $\psi\a$ and
the electromagnetic vector potential $\A$
\eqn\Hamiltonian{{F[\psi\a,\A]\over\T}=\int d^dx\left[|(\vec\nabla-i
q_o\A)\psi\a|^2+r_o|\psi\a|^2 +{1\over2} u_o (|\psi\a|^2)^2+
{1\over8\pi\mu_o}(\vec\nabla\times\A)^2\right]\;,}
where $r_o\sim(T-T_c)/T_c$, $q_o=2e/\hbar c$, and $\mu_o$ is magnetic
permeability of
the normal metal. The choice of the Coulomb gauge
$\vec\nabla\cdot\A=0$, leads to few simplifications.

I study the critical behavior of the NS transition within the self-consistent
screening approximation (SCSA) that has previously been quite successfully
applied
to a variety of other problems. \refs{\SCSAmethod}
The approximation builds on the $1/m$-expansion for general
dimensionality $d$.
\nref\Largem{J.~Zinn-Justin {\it Quantum Field Theory and Critical Phenomena}
(NY, 1989).}\refs{\Largem}
One writes downs the large $m$ limit expressions for the
renormalized interactions and propagators in terms of the bare ones and then
replaces all the bare quantities by the renormalized ones thereby obtaining
the large $m$ limit of Dyson equations for the renormalized interactions and
propagators. The advantage of this method is that in the limit
$m\rightarrow\infty$ it reduces to the exact $1/m$ result. Furthermore, while
the
straight $1/m$-expansion diverges for $m\rightarrow 0$ and therefore cannot be
taken seriously quantitatively for the physical value of $m=1$, the SCSA is
perfectly well behaved in this limit and is therefore quantitatively more
trustworthy for real superconductors.

To simplify the analysis it is convenient to integrate
out the gauge field, which can formally be done exactly since $\A$
appears at most quadratically Eq.\Hamiltonian\ , giving
$F_{tot}/\T=\int_x\left[|\vec\nabla\psi\a|^2+r_o|\psi\a|^2
+{1\over2} u_o (|\psi\a|^2)^2
-{1\over2}J_i(x) D_{ij}(x) J_j(x)+{1\over2}Tr\log(D_{ij}^{-1})\right]$.
%
%
The last two terms are the long-range effective current-current interaction
and the functional determinant generated from integration over $\A$,
respectively.
$J_i(x)=i q_o(\psi^*\a\nabla_i\psi\a-\psi\a\nabla_i\psi^*\a)$
is the paramagnetic current,
$D_{ij}^{-1}=\left[{-\nabla^2/(4\pi\mu_o})+
2 q_o^2\psi^*\a\psi\a\right]P^T_{i j} $
is the inverse of the gauge field propagator and can be read off from
Eq.\Hamiltonian\ .
%
%
The treatment of type-I superconductors is relatively simple
because the relevant temperature range lies well outside the critical region,
and therefore the order parameter fluctuations can be ignored.\refs{\HLM}
By minimizing the effective free energy
in the standard way \refs{\Largem} I obtain a gauge field corrected
mean-field theory describing the first-order NS transition previously found in
Ref.\HLM\ .

A more interesting and challenging regime is that of the type-II
superconductors where the fluctuations in the order parameter field are
strong and must be carefully taken into account. To treat this case I expand
the free energy functional in powers of $\psi\a$ to quartic order, and
because of
the smallness of the order parameter near the NS transition I ignore the
higher order corrections.
I thereby obtain an effective field theory in terms of $\psi\a$ alone,
with long-range self-interactions, described by an effective free
energy $F_{eff}[\psi\a]$
\eqn\HamiltonianEff{{F_{eff}[\psi\a]\over\T}=
\int_k\psi\a^*(k)(k^2+r_o)\psi\a(k)+
{1\over2}\int_{k_1,k_2,p}U_o(k_1,k_2,p)\psi\a^*(k_1-p)
\psi\a(k_1)\psi\b^*(k_2+p)\psi\b(k_2)\;,}
expressed in Fourier space with
$\psi\a(k)=\int d^dx\psi\a(x)e^{-i k x}$ and $\int_k=\int d^dk/(2\pi)^d$. The
effective long-range vertex of the quartic interaction is
$U_o(k_1,k_2,p)=u_o-f_o {k_{1i} k_{2j} P^T_{ij}(p)/p^2}$,
where $f_o=16\pi\mu_o q_o^2$ is the bare effective charge and
$P^T_{ij}(p)=\delta_{ij}-p_i p_j/p^2$.

Using this effective free energy I write down the coupled Dyson equations
for the renormalized $\psi\a$ propagator $G(k)$ and the renormalized quartic
interactions $u(p)$ and $f(p)$
\eqna\DysonEqns$$\eqalignno{G^{-1}(k)&=G^{-1}_o(k)+\int_p
U(k,k-p,p)G(k-p)\;,&\DysonEqns a\cr
u(p)&={u_o\over1+u_o \Pi_u(p)}\;\;,\;\;\;\;\;
f(p)={f_o\over1+f_o \Pi_f(p)}\;,&\DysonEqns b\cr}$$
%
%
%
where $\Pi_u(p)=m\int_{p'} G(p')G(p-p')$ and
$\Pi_f(p)=-m P^T_{i j}(p)/(d-1)/p^2\int_{p'} p'_i p'_j G(p')G(p-p')$
are the polarization bubbles.
The diagrammatic version of these equations is displayed in
\fig\Loops{Graphical representation of SCSA for renormalized
propagator and interaction.}.
I look for the long-wavelength-limit solutions of the above integral
equations for $G(k)$, $u(p)$ and $f(p)$. In general this can be done
numerically with the simplification that near a critical point there are
only two relevant length scales, $k^{-1}$ and the correlation length $\xi$,
and therefore for example $G^{-1}(k)=k_c^\eta k^{2-\eta}g(k\xi)$, with $g(x)$
being the scaling function and $k_c$ is a constant that depends on the
microscopics of the model.
However, exactly at criticality, $r=0$, the correlation length $\xi$
diverges and the scaling function $g(x\rightarrow\infty)=1$.
In this case $k^{-1}$ is the only relevant length scale with correlation
functions assuming even a simpler scaling form. In particular
$G^{-1}(k)=k_c^\eta k^{2-\eta}$, integral equations above can be solved
{\it exactly}, and $\eta$ determined analytically, as I demonstrate below.

Substituting the simplified scaling form for $G^{-1}(k)$ into $\Pi_u(p)$ and
$\Pi_f(p)$ I find,
%
%
%
\eqn\Piesii{\Pi_u(p)=m
I_0(1-\eta/2,1-\eta/2)k_c^{-2\eta}p^{d-4+2\eta}\;,\;\;\;
\Pi_f(p)=m
I_{02}(1-\eta/2,1-\eta/2)k_c^{-2\eta}p^{d-4+2\eta}\;,}
where I defined integrals,
$I_0(a,b)=\int_{p'} ({\hat p}-p')^{-2a}/p'^{-2b}=
\Gamma(a+b-d/2)\Gamma(d/2-a)\Gamma(d/2-b)/(4\pi)^{d/2}
/\Gamma(a)/\Gamma(b)/\Gamma(d-a-b)$, $I_{ij}(a,b)
=\int_{p'} p'_i p'_j ({\hat p}-p')^{-2a} p'^{-2b}
=\delta_{ij}I_{02}(a,b)+ {\hat p}_i {\hat p}_j I_{22}(a,b)$,
$I_{02}(a,b)=-\Gamma(a+b-d/2-1)\Gamma(d/2-a+1)\Gamma(d/2-b+1)/
2/(4\pi)^{d/2}/\Gamma(a)/\Gamma(b)/\Gamma(d-a-b+2)$,
$I_{22}(a,b)=\Gamma(a+b-d/2)\Gamma(d/2-a)\Gamma(d/2-b+2)
/(4\pi)^{d/2}/\Gamma(a)/\Gamma(b)/\Gamma(d-a-b+2)$,
%
%
and $\hat p$ is a unit vector.

I first look at the Heisenberg critical point by setting $f_o=0$, which
automatically leads to $f(p)=0$ from Eq.\DysonEqns{b}. Assuming that
$4-d>2\eta$, (this assumption will be satisfied by the solution for $\eta$ for
$d<4$; for $d>4$ the Gaussian fixed point result is recovered)
in the long wavelength limit, $p\rightarrow 0$, $\Pi_u(p)$
dominates over the $1$ in the denominator of Eq.\DysonEqns{b} and
the renormalized $u(p)$ interaction reduces to a universal function,
\eqn\uH{u(p)=\Pi^{-1}_u(p)
={1\over mI_0(1-\eta/2,1-\eta/2)}k_c^{2\eta}p^{-d+4-2\eta}\;.}
Using the renormalized version of the quartic interaction together with above
equations and the scaling form for $G(k)$ in Eq.\DysonEqns{a}\ , I find that
$\eta$ is determined by $m=I_0(1-\eta/2,\eta+d/2-2)/I_0(1-\eta/2,1-\eta/2)$.
%
%
For the physical superfluids with $m=1$ and $d=3$ this result leads to
$\eta\approx 0.125$.
The implicit equation for $\eta(d,m)$ can also be expanded
in $\eps=4-d$ or $1/m$ in which case I obtain
$\eta_\eps=\eps^2/(4m)$ (arbitrary $m$) and $\eta_m=4/(3\pi^2 m)$ ($d=3$),
respectively. The large $m$ limit of this result, $\eta_m$,
by construction agrees
exactly with the direct $1/m$-expansion result. However, the
$d\rightarrow 4$ limit of SCSA, $\eta_\eps$,
does not get the $m$ dependence quite correctly when compared to the leading
order in the $\eps$-expansion, where the result is $\eta=\eps^2(1+m)/(8+2m)^2$.
I note that this disagreement with $\eps$-expansion for small $m$ is expected
from the fact that SCSA does not correctly account for triple vertex
renormalization which are taken into account in the direct $\eps$-expansion for
arbitrary $m$.\refs{\Largem}

I now apply the above calculations to the full problem of the NS transition.
Allowing now for a nonzero charge, $f_o\neq 0$, I arrive at the charged
analogs of
Eqs.\uH. The expression for $u(p)$ is the same, and $f(p)$ in the
asymptotic limit reduces to
\eqn\fH{f(p)=\Pi^{-1}_f(p)
={1\over mI_{02}(1-\eta/2,1-\eta/2)}k_c^{2\eta}p^{-d+4-2\eta}\;.}
Substituting these screened interactions into $U$, defined by the
renormalized version of $U_0$ and then using
Eq.\DysonEqns{a} I obtain,
\eqn\Dysonii{k_c^\eta k^{2-\eta}={k_c^\eta\over
m}\int_p\left({I_0^{-1}(1-\eta/2,1-\eta/2)\over(k-p)^{2-\eta}p^{d-4+2\eta}}-
{I_{02}^{-1}(1-\eta/2,1-\eta/2)k_i k_j P^T_{i j}(p)\over(k-p)^{2-\eta}
p^{d-2+2\eta}}\right)\;.}
Performing above integrals leads to the equation which determines
$\eta(d,m)$ at the new superconducting critical point,
\eqnn\etaSC$$\eqalignno{m=&
{I_0(1-\eta/2,\eta+d/2-2)\over I_0(1-\eta/2,1-\eta/2)}&\cr
-&{I_0(1-\eta/2,\eta+d/2-1)+I_{02}(1-\eta/2,\eta+d/2)-
I_{22}(1-\eta/2,\eta+d/2)\over I_{02}(1-\eta/2,1-\eta/2)}\;.&\etaSC\cr}$$

The above implicit result for $\eta(d,m)$ reduces to
$\eta_\eps=-9\eps/(m-18)$ in the
limit of $d\rightarrow 4$, for arbitrary $m$. In the regime where HLM find
the NS critical fixed point ($m>183$), this $\eps$-expansion result is less
than within 10\% of their exact (to $O(\eps)$ ) value of
$\eta^{HLM}_\eps=-9\eps/m$.\refs{\HLM} \
It is important to note that although the complete SCSA result
for $\eta$ (Eq.\etaSC) is well behaved as a function of $m$
(see \fig\Graph{$\eta^{SCSA}$ (for $d=3$, full curve)
plotted as a function of $m$, showing
improvement at small $m$ compared to the direct $1/m$-expansion result,
$\eta_m=-20/(\pi^2 m)$ (dashed curve).
The inset shows $\eta^{SCSA}$ as a function of $m$ for
various values of $d$.}),
it breaks down at a
critical value of $m_c=18$, when expanded in $\eps$. This suggests
that the dissappearance of the critical point and the runaway rg flows
for $m<m_c\approx 183$ in the direct $\eps$-expansion of Ref.\HLM\ \
should be interpreted as the breakdown of the $\epsilon$-expansion rather
than the fluctuation driven first-order transition.

Expanding the result, Eq.\etaSC\ , for large $m$ in powers of $1/m$ I
recover the results of HLM in this limit, obtaining $\eta_m=-20/\pi^2
m\approx -2.026/m$, for $d=3$. It is important to
note, however, that the direct $1/m$-expansion leads to the value
of $\eta_m=-2.026$ (for $d=3, m=1$) that lies outside the physical range
$\eta>2-d=-1$. In contrast the SCSA approximation gives a sensible result of
$\eta=-0.38$ for real superconductors, that is well within this physical
range.  For $d>4$, I recover the Gaussian fixed point,
as expected since the upper-critical dimension for the NS transition is
$d_{uc}=4$. The SCSA then serves as a physical interpolation between
large $m$ and small $\eps$ behavior.

In conclusion, I find that the  self-consistently modified large $m$-
expansion for the generalized theory of the NS transition leads to a
nontrivial critical point for 3d superconductors.
In contrast to the original HLM interpretation I
predict a second-order transition for type-II superconductors in $d=3$,
in qualitative agreement with the work of Dasgupta and Halperin and with
the related theoretical and experimental findings for the NA
transition. The results suggest a break down of the
$\epsilon$-expansion below a critical value of $m$,
while the actual NS transition remains continuous.

This research was supported by the NSF (DMR $91-20000$)
through the STCS. It is a pleasure to acknowledge discussions with
David Land that stimulated my interest in this problem. I thank Professors
Bert Halperin, David Nelson and Tom Lubensky for helpful
conversation and an anonymous referee for valuable suggestions,
references, and humor.
\vfill\eject

\vfill\eject
\listrefs
\vfill\eject
\listfigs
\vfill\eject
\bye